\documentclass[12pt,preprint]{aastex}
\usepackage{natbib}
\usepackage{graphicx}
\newcommand{\eqb}{\begin{equation}}
\newcommand{\eqe}{\end{equation}}

\begin{document}

\title{Adjustment of the electric charge and current in pulsar magnetospheres}

\author{Yuri Lyubarsky}
 \affil{Physics Department, Ben-Gurion University, P.O.B. 653, Beer-Sheva 84105, Israel}

\begin{abstract}
We present a simple numerical model of the plasma flow within the open field line tube in the pulsar magnetosphere. We study how the plasma screens the rotationally induced electric field and maintains the electric current demanded by the global structure of the magnetosphere. We show that even though bulk of the plasma moves outwards with relativistic velocities, a small fraction of particles is continuously redirected back forming reverse plasma flows. The density and composition (positrons or electrons, or both) of these reverse flows are determined by the distribution of the Goldreich-Julian charge density along the tube and by the global magnetospheric current. These reverse flows could significantly affect the process of the pair plasma production in the polar cap accelerator. Our simulations also show that  formation of the reverse flows is accompanied by the generation of long wavelength plasma oscillations, which could be converted, via the induced scattering on the bulk plasma flow, into the observed radio emission.
\end{abstract}


\maketitle
\section{Introduction}
The pulsar activity is believed to be associated with the
generation of relativistic electron-positron plasma near the
magnetic polar caps.
This plasma flows along the open field line tube and eventually
escapes from the magnetosphere forming a relativistic pulsar
wind. Well within the light cylinder, the plasma currents do
not distort significantly star's magnetic field therefore in
the frame corotating with the star, the plasma just moves along
the axis of the rotating static dipole. An important point is
that this motion could by no means be considered as free
streaming because the basic electrodynamics dictates the charge
and current densities at each point of the flow.

First of all the charge density in the plasma should be equal
to the local Goldreich-Julian charge density
 \eqb
\rho_{\rm GJ}=-\frac{\mathbf{\Omega\cdot B}}{2\pi c};
 \label{GJ}\eqe
where $B$ is the local magnetic field, $\Omega$ the angular
velocity of the neutron star. This condition is a
generalization of the standard condition of the
quasi-neutrality: deviation of the local charge density
from  $\rho_{\rm GJ}$  results in a longitudinal electric field, which
redistributes the charges to establish $\rho=\rho_{\rm GJ}$. The
necessary field is weak in the sense that the corresponding
potential is small as compared with the total rotationally
induced potential; this is because the energy of the secondary
plasma particles is small as compared with the total potential.

   \citet{scharlemann74} and \citet{cheng_rud77} noted that in the
plasma flow along the curved magnetic field lines, some
difference should be maintained between the velocities of the
electrons and positrons. The condition $\rho=\rho_{\rm GJ}$
implies $n^+-n^-\propto B\cos\theta$, where $n^{\pm}$  are the
number density of positrons and electrons, correspondingly,
$\theta$ the angle between the pulsar rotation axis and the
magnetic field. Continuity of the particle flow implies
$n^{\pm}V^{\pm}\propto B$, where $V^{\pm}$  are the average
velocities of the positrons and electrons, correspondingly. One
sees that the average particle velocities should vary along a
curved magnetic field line in accordance with the variation of
$\theta$. In a highly relativistic flow, $V^{\pm}$ are close to
$c$ therefore even a small variation of $\theta$ implies a
significant variation of the Lorentz factor of the positrons
and electrons. Since the energy spread of the secondary
particles is large, the electric field, which ensures the
adjustment of the charge density, easily shifts the low-energy
tail of the distribution function to the negative-momentum
domain thereby forming a return particle flux
\citep{lyub_current92, lyub_turb93}. Therefore one can expect
counterstreaming plasma flows in the open field line tube.

    If the charge density in the flow is adjusted via sending "extra"
charges backwards, an electric current appears in the flow.
However, the electric current is not a free parameter, which
could be adjusted to maintain the necessary distribution of
local charges. The current in the open field line tube is
dictated by the global structure of the magnetosphere. The
magnetic field lines are bent backwards with respect to the
rotation direction in order to allow the plasma to stream along
them at a velocity smaller than the speed of light even though
the rotation velocity becomes superluminal beyond the light
cylinder. The current should be distributed such that the
necessary magnetic field is maintained (formally, the current
along any magnetic field line is determined from the condition
of the smooth transition of the flow through the light cylinder
\citep{ingraham73, contopoulos99,timokhin06}).
Generally the current required by the global structure of the
magnetosphere is not matched with the current established in
the course of the charge density adjustment. This would result
in some induction electric field, which ensues additional
redistribution of the charge flows. These considerations show
that the self-consistent distribution of charges and currents
in the open field line tube could be maintained only in a
non-trivial, generally unsteady, multi-stream state.

As the first step to understanding of the electric charge and
current adjustment in the pulsar magnetospheres, we study here
a simple model of the plasma motion within a narrow tube with a
distributed background charge imitating the Goldreich-Julian
charge. The plasma production is not addressed here; we just
assume that the relativistic electron-positron plasma with
large enough density (significantly larger than the
Goldreich-Julian density) is injected continuously into the
tube. The aim of this study is to keep track of how the plasma
screens out the background charge and at the same time
maintains the current imposed at the outer edge of the tube.
The system evolution is studied numerically by the
particle-in-the-cell (PIC) method.

It will be shown that the electric charge and the current
densities are adjusted in the plasma flow such that the
condition $\rho=\rho_{\rm GJ}$ is fulfilled at any point and at
the same time the current imposed by the boundary conditions is
maintained. This justifies the MHD approach in studying the
global structure of the pulsar magnetosphere. The charge and
current adjustment is achieved via formation of relatively weak
backward flows. This could have important implications for the
models of the plasma production in pulsars because the plasma
flow from the magnetosphere onto the polar cap could alter
significantly the electric structure of the polar gap
accelerator. Moreover, long wave plasma oscillations are
excited in the flow in the process of continuous redistribution
of charges. The induced scattering of these oscillations by the
bulk plasma flow could give rise to the observed pulsar radio
emission.

The paper is organized as follows. In the next section, the
numerical method is outlined. In section 3, we present the
results of PIC simulations of the relativistic plasma flow
within a narrow tube with a distributed background charge. In
section 4, we discuss the plasma oscillations observed in the
simulations. In sect. 5, we generalize the model adding an
arbitrary current at the outer edge of the tube. The
conclusions are given in section 6.

\section{Quasi-1D electrodynamics}
In this section, a simple generalization of the one-dimensional
PIC code is described, which pinpoints the basic qualitative
properties of the plasma motion within the open field line
tube. Since the particles move along the magnetic field lines like
beads on a wire, their motion is governed only by the
longitudinal electric field,
 \eqb
\frac{dp}{dt}=\pm eE_z.
 \eqe
 Therefore one can use the particle
mover from the electrostatic one-dimensional PIC code (e.g.,
\citet{birdsall_langdon}). The problem is that in our case, the
electric field is not one-dimensional because the plasma moves
within a narrow tube; therefore one has to solve Maxwell's
equations within the tube. We take into account the effects of
finite transverse dimensions by substituting $1/R$, where $R$
is the tube radius, for the transverse derivatives in Maxwell's
equations. The same substitution is sometimes made in the
Poisson equation when studying steady state flows in the open
field line tube \citep{fawley77}.

Maxwell's equations in the frame rotating with the star could
be written as customary Maxwell's equations with specific
source terms (e.g., \citet{fawley77})
 \begin{eqnarray}
& & \mathbf{\nabla\cdot E}=4\pi (\rho-\widetilde{\rho});
 \label{Gauss} \\
& & \mathbf{\nabla\cdot B}=0; \label{B} \\
& & \mathbf{\nabla\times B}=\frac{4\pi}c
(\mathbf{j}-\widetilde{\mathbf{j}})+\frac{\partial \mathbf{E}}{\partial
t}; \label{Ampere} \\
& & \mathbf{\nabla\times E}=-\frac{\partial \mathbf{B}}{\partial t}.
 \label{Faraday}
 \end{eqnarray}
Well inside the light cylinder, $\Omega r\ll 1$, one can take
$\widetilde{\rho}=\rho_{\rm GJ}$; $\widetilde{\mathbf{j}}=0$.

Evolution of the fields is governed by Eqs.(\ref{Ampere}) and
(\ref{Faraday}); Eqs. (\ref{Gauss}) and (\ref{B}) are satisfied
identically provided the initial conditions satisfy these
equations and the charge is conserved. Within the cylindrical
tube, the evolution equations are written as
\begin{eqnarray}
& & \frac{\partial E_z}{\partial t} = \frac 1r\frac{\partial rB_{\varphi}}{\partial r} -
\frac{4\pi}c j; \nonumber  \\
& & \frac{\partial E_r}{\partial t} = -\frac{\partial B_{\varphi}}{\partial z}; \label{Maxwell}\\
&& \frac{\partial B_{\varphi}}{\partial t} = -\frac{\partial E_r}{\partial z} +
 \frac{\partial E_z}{\partial r}.\nonumber
 \end{eqnarray}
Here $B_{\varphi}$ is the azimuthal magnetic field created by
the currents in the plasma. Well within the light cylinder, this field is
very weak as compared with the pulsar magnetic field therefore it does not
affect the particle motion; the particles move in the $z$ direction along
the pulsar magnetic field lines. 
We assume that the walls of the tube are highly conductive so
that $E_z=0$ at the boundary of the tube, $r=R$. Substituting the
transverse derivatives by $1/R$, we get a
one-dimensional set of equations
\begin{eqnarray}
& & \frac{\partial E_z}{\partial t} =\frac{B_{\varphi}}R - \frac{4\pi}c j; \nonumber\\
& & \frac{\partial E_r}{\partial t} = -\frac{\partial B_{\varphi}}{\partial z}; \label{mainEq}\\
& & \frac{\partial B_{\varphi}}{\partial t} = -\frac{\partial E_r}{\partial
z} - \frac{E_z}R; \nonumber\end{eqnarray}
which is qualitatively equivalent to the original Eqs. (\ref{Maxwell}).
 Here we took into account that $E_z$ goes to zero at the wall so that
$\partial E_z/\partial r\Rightarrow -E_z/R$. Eqs. (\ref{mainEq}) could be reduced to the one-dimensional,
inhomogeneous Klein-Gordon equation. At the scale much less
than the radius of the tube, these equations are reduced to the
one-dimensional Maxwell equations. At larger scales, they
correctly describe the exponential decreasing of the electric
field from a localized charge due to the
presence of conducting walls. For boundary conditions, one sets
$E_r$ at the bottom edge of the tube and $B_{\varphi}$ at the
outer edge.

We start from the empty tube therefore $B_{\varphi}(t=0)=0$
whereas the initial electric field is found from the solution
to the Gauss equation (\ref{Gauss}) with $\rho=0$. With the
same prescription for the transverse derivatives as in
equations (\ref{mainEq}), one can write $E_r(t=0)=\Phi/R$;
$E_z(t=0)=-\partial\Phi/\partial z$; where the potential
$\Phi$ is found from the modified Poisson equation
 \eqb
\frac{d^2\Phi}{dz^2}-\frac{\Phi}{R^2}=4\pi\rho_{\rm GJ}.
 \label{Poisson}\eqe

With the initial conditions thus found, the evolution equations
(\ref{mainEq}) are solved by the leap-frog method:
\begin{eqnarray}
& & E_{z}(z,t+\Delta t) - E_{z}(z,t) = \frac{\Delta t}R B_{\varphi}(z,t+0.5\Delta t) -
\frac{4\pi}c \Delta t j(z,t+0.5\Delta t) \nonumber \\
& & E_{r}(z+0.5\Delta z,t+\Delta t) - E_{r}(z+0.5\Delta z,t)\nonumber \\
& &\qquad = -\frac{\Delta t}{\Delta z}\left[ B_{\varphi}(z+\Delta z,t+0.5\Delta t) - B_{\varphi}(z,t+0.5\Delta t) \right]; \\
& & B_{\varphi}(z,t+0.5\Delta t) - B_{\varphi}(z,t-0.5\Delta t)\nonumber \\
& & \qquad= -\frac{\Delta t}{\Delta z}\left[ E_{r}(z+0.5\Delta z,t) - E_{r}(z-0.5\Delta z,t) \right] -
 \frac{\Delta t}R  E_{z}(z,t). \nonumber
\end{eqnarray}
The current density, $j$, is found in each step by calculating
the charge transferred by particles through each grid face. We
use the first order particle weighting, i.e. the particle is
assumed to be a cloud of uniform density, the cloud width being
equal to the grid size.

Stability of the scheme is analyzed as usual by writing down
the dispersion relation for the homogeneous system:
 \eqb
\left(\frac{2R}{\Delta
t}\right)^2\sin^2\left(\frac{\omega\Delta t}2\right) = 1 +
\left(\frac{2R}{\Delta z}\right)^2 \sin^2\left(\frac{k\Delta
z}2\right).
 \label{disp1}\eqe
One sees that under the condition
 \eqb
\left( \Delta t\right)^2 \le \frac{(\Delta z)^2}{1+(\Delta
z/2R)^2}
 \label{courant}\eqe
the frequency $\omega$ is real, which implies stability. In the
simulations, we used $\Delta t=0.5\Delta z$.


\section{Screening of the rotationally induced electric field}

The rotationally induced electric field implies specific source
terms in Maxwell's equations written in the rotating frame of
reference. Deep inside the light cylinder these terms are
reduced to the Goldreich-Julian charge density  of equation
(\ref{GJ}), which could be considered as a background charge
distributed within the tube. Our main interest here is with the
compensation of the background charge in the relativistic
plasma flow. In the static system, the particles of
both charge species would easily be redistributed to provide
total neutrality. The situation is less trivial when the plasma
flows through the tube. We investigate the simplest case when the plasma
density in the flow, $n$, is large enough to compensate the
background charge, $en\gg \rho_{\rm GJ}$.

For the sake of simplicity, we assume that the radius of the
tube, $R$, is constant. In the pulsar magnetosphere, the open field
line tube expands with the altitude, which results in
decreasing of the particle density as well as of the charge and
current densities. However, the Goldreich-Julian charge density
(\ref{GJ}) decreases by the same factor because the pulsar magnetic
field decreases in the expanding tube. Therefore only variation
of the angle between the magnetic field and the rotation axis,
but not the tube expansion, results in the discrepancy between
$\rho$ and $\rho_{\rm GJ}$. Thus the plasma flow in a tube of
a constant radius with a varying background charge seems to be
a reasonable model for studying screening of the Goldreich-Julian charge in the
open field line tube. Note also that only variation of the background charge
along the tube results in the nontrivial behavior of the plasma.
If $\rho_{\rm GJ}=\it const$, the charge density would be adjusted already at the bottom of
the tube and then the acquired charge would be just transferred by the flow further into the tube. As we are
interested in a continuous charge redistribution in the plasma flow along the tube,
we take the background charge to be varying along the tube. The simplest choice is
a linearly growing background charge density
 \eqb
\rho_{\rm GJ}=\rho_0z/l_0;
 \label{shortube}\eqe
 where $l_0$ is the length of the tube.
Then the system is neutral at the bottom of the tube whereas at higher
altitudes, charge neutrality may be maintained only via
redistribution of charges in order to compensate the growing
background charge.

We took the radius of the tube much smaller than the tube
length, $R=l_0/30$, which means that different parts of the
tube are not connected electrically. The initial electric field
was found from the solution to the Poisson equation
(\ref{Poisson}) with $E_z=0$ at both edges of the tube. One can
see that far from the edges of the tube, $z\gg R$ and $l_0-z\gg
R$, the solution to equation (\ref{Poisson}) with the
background charge density (\ref{shortube}) is $\Phi=-4\pi\rho_0
R^2z/l_0$. Then the initial longitudinal electric field is
constant far from the edges of the tube, \eqb
E_z(t=0)=4\pi\rho_0R^2/l_0; \label{Ez_init}\eqe whereas the
transverse electric field increases linearly from the bottom to
the upper edge of the tube,
 \eqb
 E_r(t=0)=-4\pi\rho_0 Rz/l_0.
 \label{Er_init}\eqe

\begin{figure*}
\includegraphics[width=14 cm,scale=1]{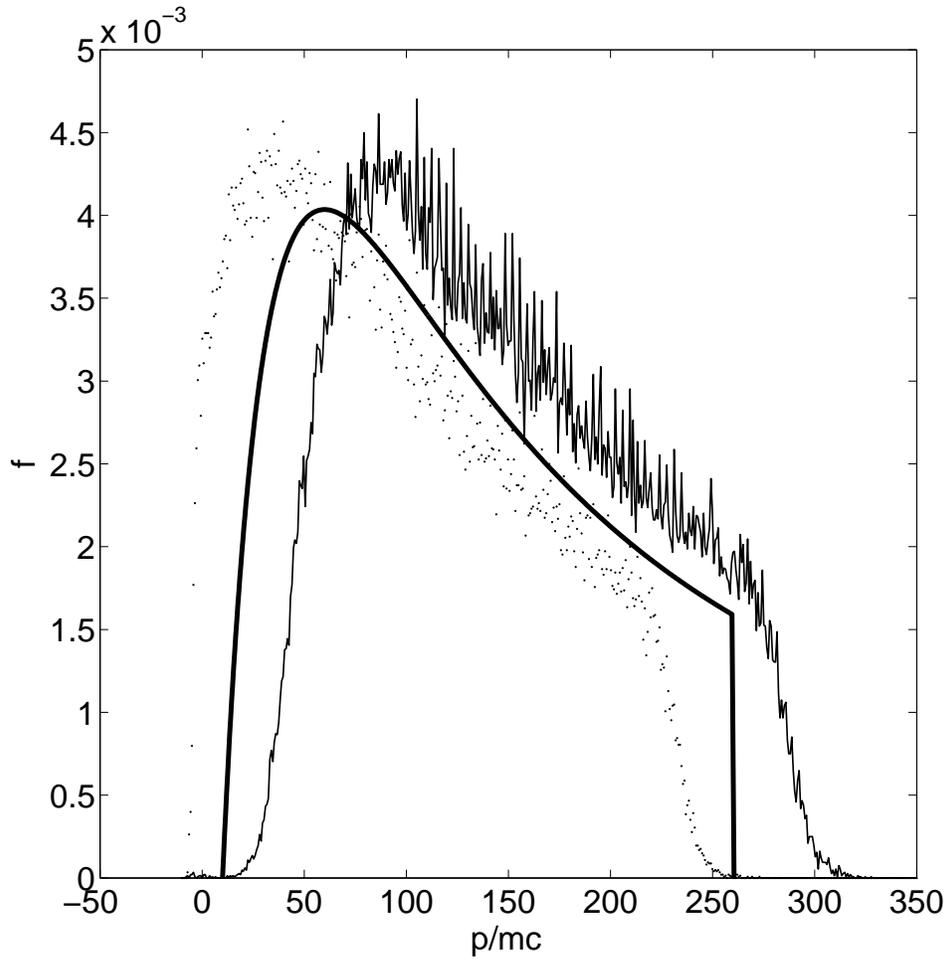}
\caption{The distribution function of the injected particles
(thick line), as well as the distribution functions of
positrons (thin line) and electrons (dots) found in the region
$0.25l_0<z<0.75l_0$ at the end of simulations, at
$t=3.25l_0/c$.}
\end{figure*}

We start injecting plasma from the bottom into the empty tube.
At the bottom boundary, we use the condition $E_r(z=0)=0$,
which follows from the continuity of the transverse component
of the electric field. At the upper boundary, we take
$B_{\varphi}(z=l_0)=0$. Then the plasma moves as within an
infinitely long tube until the flow approaches the outer edge
of the tube by the distance $R$, i.e. while the plasma is
disconnected electrically from the outer boundary. Electrons
and positrons are injected continuously with the same density
and distribution function; the last is shown in Fig. 1. The
pair density in the flow is $n=10$ pairs per cell. We choose
$\rho_0=0.1en$ so that the screening of the background charge
would require redistribution of a small fraction of the
particles. The electron charge, $e$, was chosen such that
$\omega_p\Delta z/c=0.3$ whereas the radius of the tube was
$R=1000\Delta z$. Introducing is the characteristic plasma
frequency, $\omega_p=\sqrt{8\pi e^2n/m}$ (recall that $n$ is
the number density of pairs therefore the total particle number
density is $2n$), one sees that $\omega_pR/c=300$. With the
chosen parameters, the total potential drop in the empty tube
is
 \eqb
e\Phi_0=4\pi e\rho_0R^2=\frac{\rho_0}{2en}
m\omega_p^2R^2=4500mc^2;
 \eqe
  which is much larger than the
energy of the injected particles. Therefore the first injected
particles are accelerated to high energies. However within the
plasma, the electric field is nearly shorted out. Compensation
of the background charge is provided by decelerating and
sending back "extra" charges (electrons in our case) therefore
within the plasma flow, a small electric field is maintained
sufficient to redirect the slowest particles.

Evolution of the fields and the particle phase space
distribution is shown in Figs. 2 and 3. One sees that an
electron backflow is indeed formed and the plasma efficiently
screens the longitudinal electric field. At the bottom of the
tube, a relatively large electric field arises with the total
potential drop roughly $10 mc^2/e$. This is because in the
injected plasma, there are no particles with the energy less
than $10mc^2$  so that a "double layer" is formed in order to
shift the low energy tail of the electron distribution to the
negative momentum region. In the remainder part of the tube,
the field is very low so that the total potential drop is
comparable with that near the bottom edge. The distribution of
the electric potential along the tube is determined by the
distribution of the background charge and by the particle
distribution function in the low energy part. The background
charge dictates how many electrons should be removed from the
flow  and the potential is adjusted such that the necessary
amount of electrons is redirected back.

As the plasma moves farther into the tube, the increasing
number of electrons is sent back in order to compensate the
increasing background charge. The increasing reverse flux of
electrons implies the increasing electric current flowing in
the plasma and therefore the increasing azimuthal magnetic
field. Note that the plasma screens  immediately only the
longitudinal electric field whereas the transverse electric
field remains nonzero; it only decreases as compared with the
initial transverse field. As the transverse field varies along
the tube, the curl of the electric field arises, which is
balanced, according to the induction law, by the growing
magnetic field.

\begin{figure*}
\includegraphics[width=14 cm,scale=1]{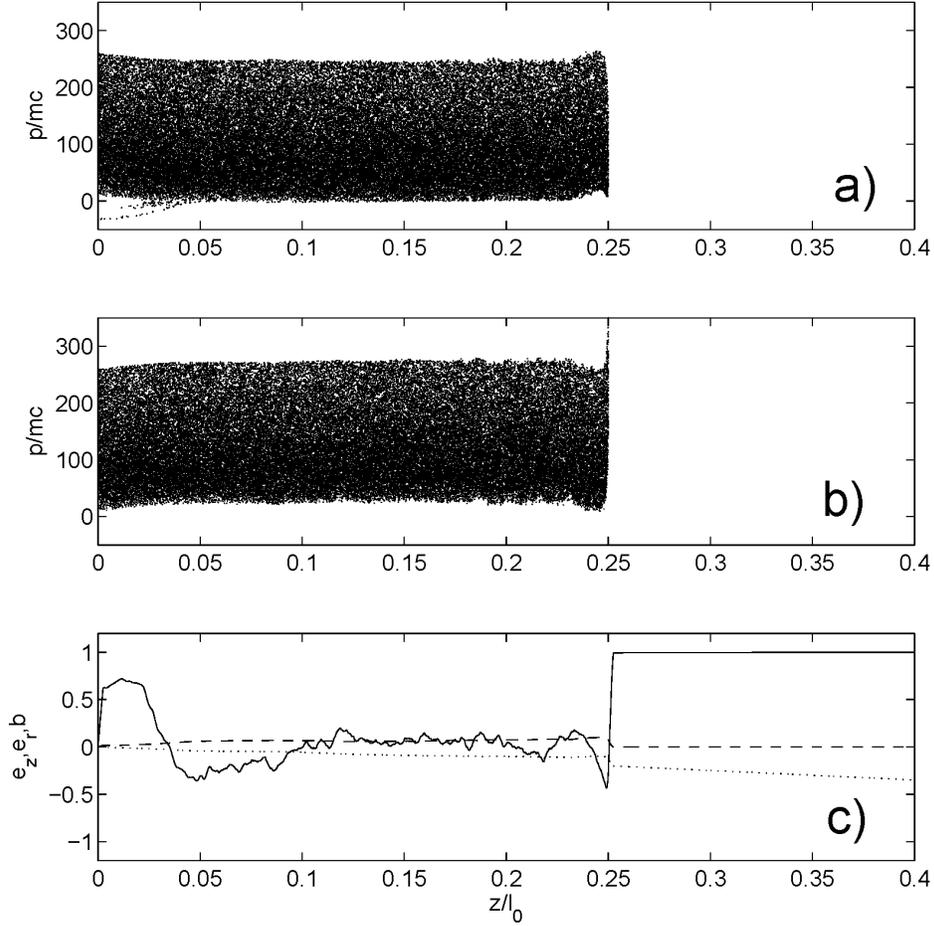}
\caption{The phase space of electrons (a), positrons (b) and
the distribution of the normalized fields,
$e_z=E_zl_0/(4\pi\rho_0 R^2)$ (solid), $e_r=E_r/(4\pi\rho_0R)$
(dotted), $b=B_{\varphi}/(4\pi\rho_0 R)$ (dashed), at
$t=0.25l_0/c$. The fields are smoothed at a scale $2\pi
c/\omega_p$.}
\end{figure*}

\begin{figure*}
\includegraphics[width=14 cm,scale=1]{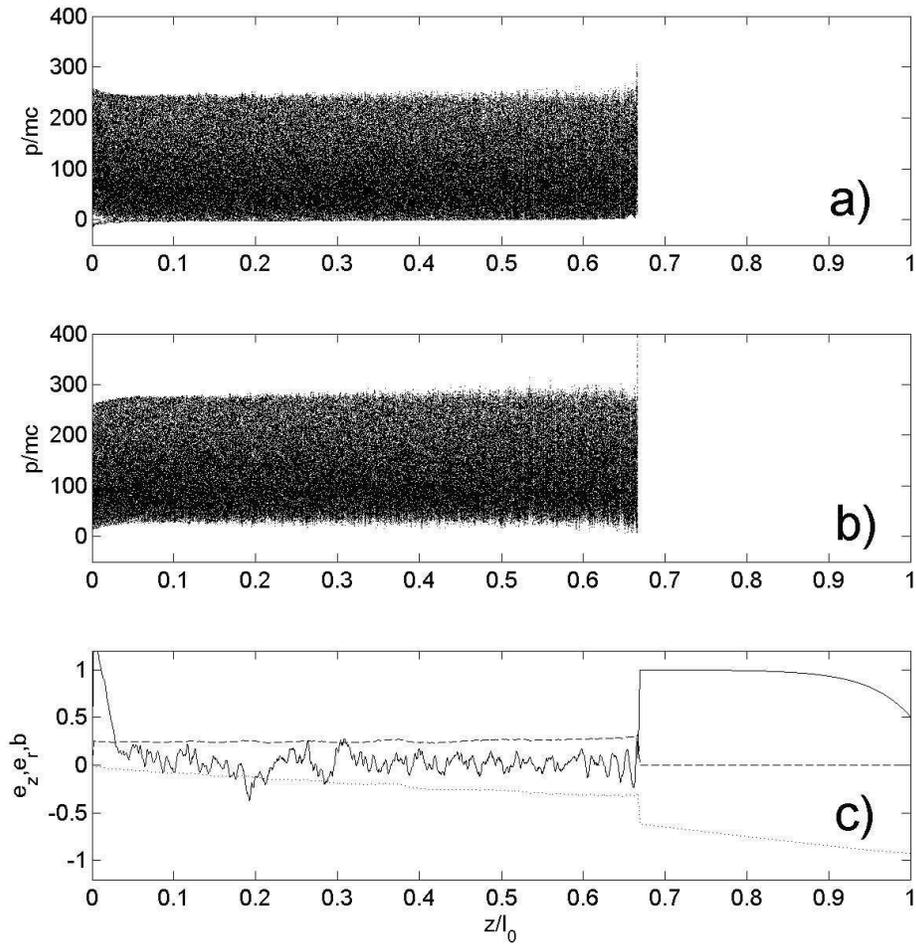}
\caption{The same as in Fig. 2 at $t=0.67l_0/c$.}
\end{figure*}

\begin{figure*}
\includegraphics[width=14 cm,scale=1]{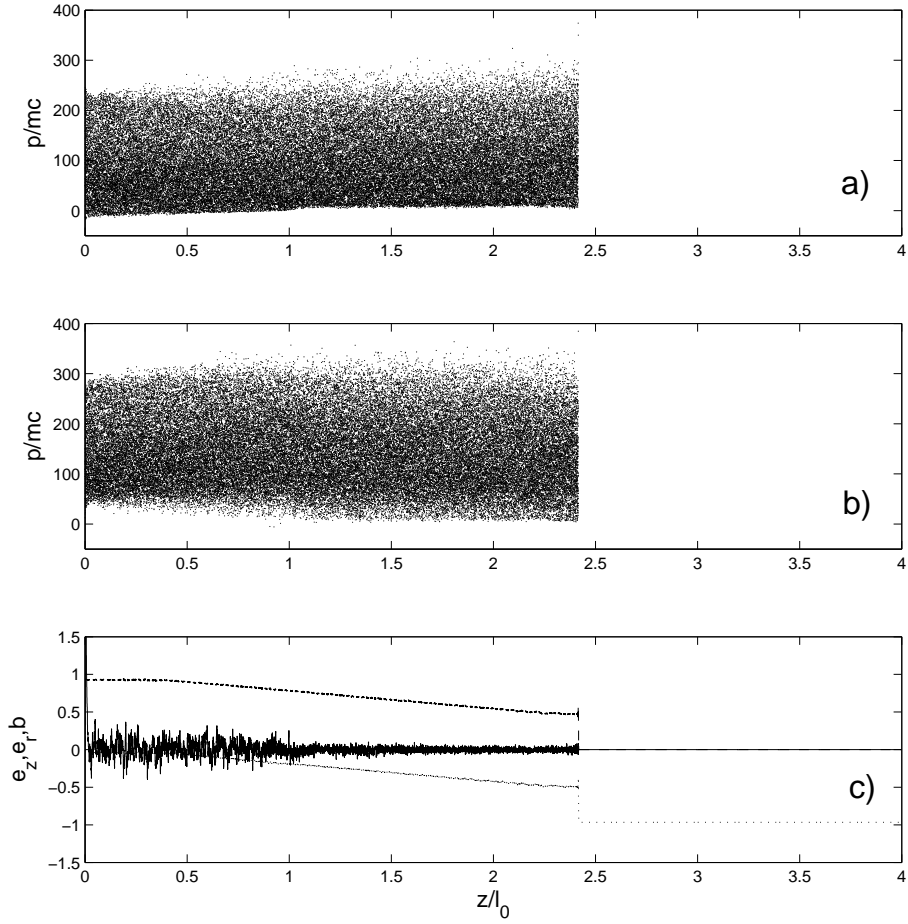}
\caption{The same as in Fig.2 at $t=2.4l_0/c$.}
\end{figure*}

\begin{figure*}
\includegraphics[width=14 cm,scale=1]{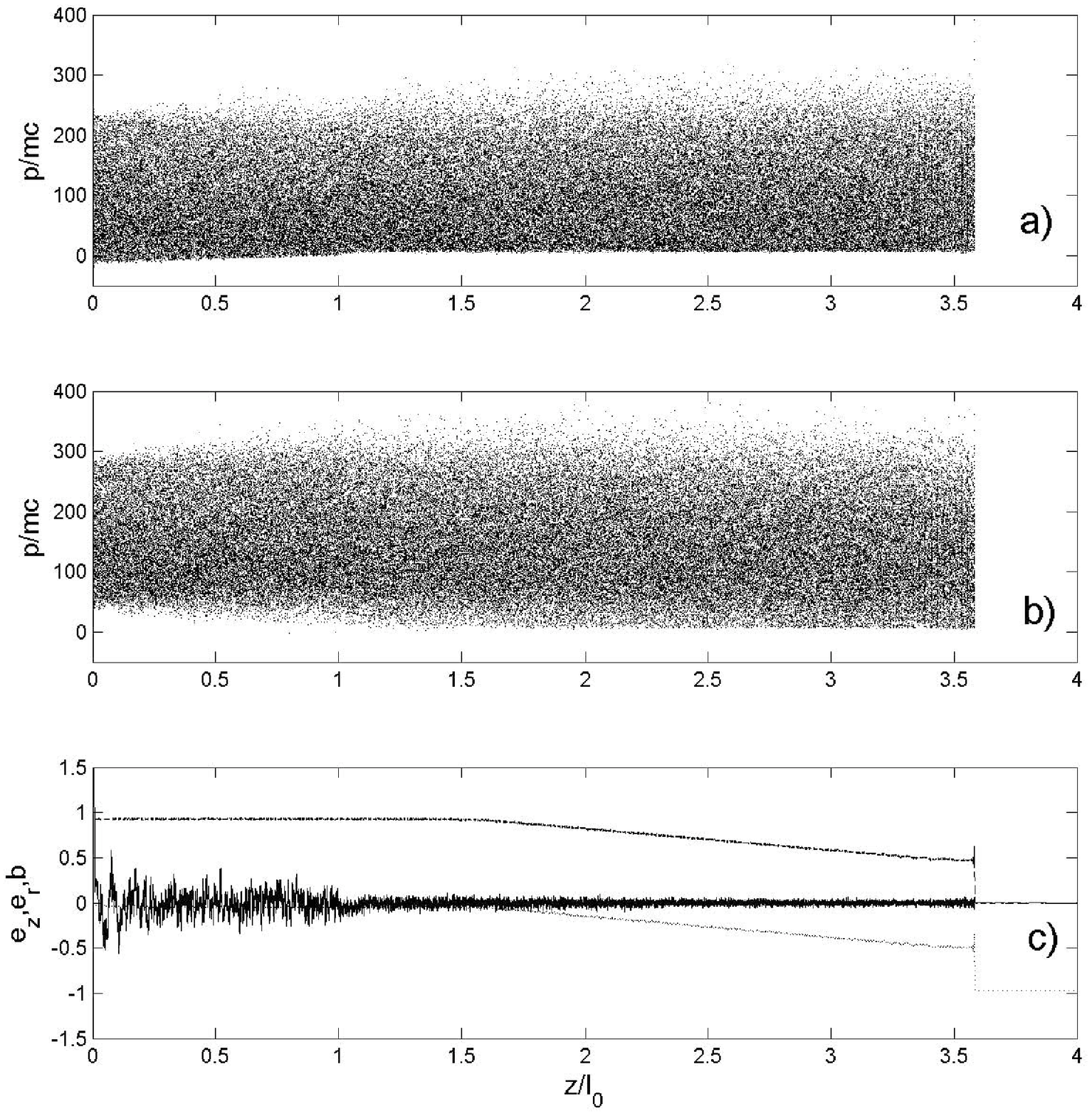}
\caption{The same as in Fig. 2 at $t=3.6l_0/c$. The steady
state is formed at $z<l_0$.}
\end{figure*}

In order to check whether a steady state solution is possible,
at least in the average, we extend the tube to $z>l_0$ assuming
that the background charge density remains constant at $z>l_0$:
 \eqb
\rho_{\rm GJ}=\left\{\begin{array}{ll} \rho_0 z/l_0; & z<l_0;
\\ \rho_0; & z>l_0.\end{array}\right.
\label{longtube} \eqe
 One can anticipate that in the region
with $\rho_{\rm GJ}=\it const$, the reverse plasma flow is not
formed therefore this region would not affect the region
$z<l_0$ so that a steady state solution would be eventually
formed there. The simulations show that this is indeed the case
(see Figs. 4 and 5). The transverse field gradually decreases
to zero and a steady state is formed with ${\mathbf{E}}=0$ and
$\rho=\rho_{\rm GJ}$. In the course of time, the steady zone extends further into the tube.

One can easily estimate the current, and therefore the
azimuthal magnetic field, established in the steady state. In the region $z>l_0$,
the background charge density is constant, $\rho_{\rm GJ}=\rho_0$;
the particle redistribution does not occur there and the reverse particle flux
 is not formed. Then the current may be presented as $j=ec(n_+-n_-)$, where $n_+$
and $n_-$ are the number densities of positrons and electrons,
correspondingly. Here we assume that all particles move  with
the speed of light. In the pulsar magnetospheres, the particle
distribution is wide, $\Delta\gamma\sim\gamma\sim 100$, and in
the presence of the longitudinal electric field, the
distribution function is just shifted in the momentum space  as
a whole  so that
the fraction of the particles with the modulo velocities
significantly less than the speed of light remains negligibly small in any
case. Taking into account that charge neutrality implies
$e(n_+-n_-)=\rho_0$, one finds finally $j=c\rho_0$. When the steady state is achieved,
the magnetic field approaches
 \eqb
B_{\varphi,0}=4\pi R\rho_0.
 \label{B0}\eqe
This estimate assumes that the electric field in the tube is
screened completely. However, some small field remains in order
to decelerate the necessary amount of electrons and send them
backwards. Therefore the charge density in the plasma is a bit
less than the background charge density, which implies that the
magnetic field remains a bit less than that of equation
(\ref{B0}); see Fig. 5.

\begin{figure*}
\includegraphics[scale=0.45]{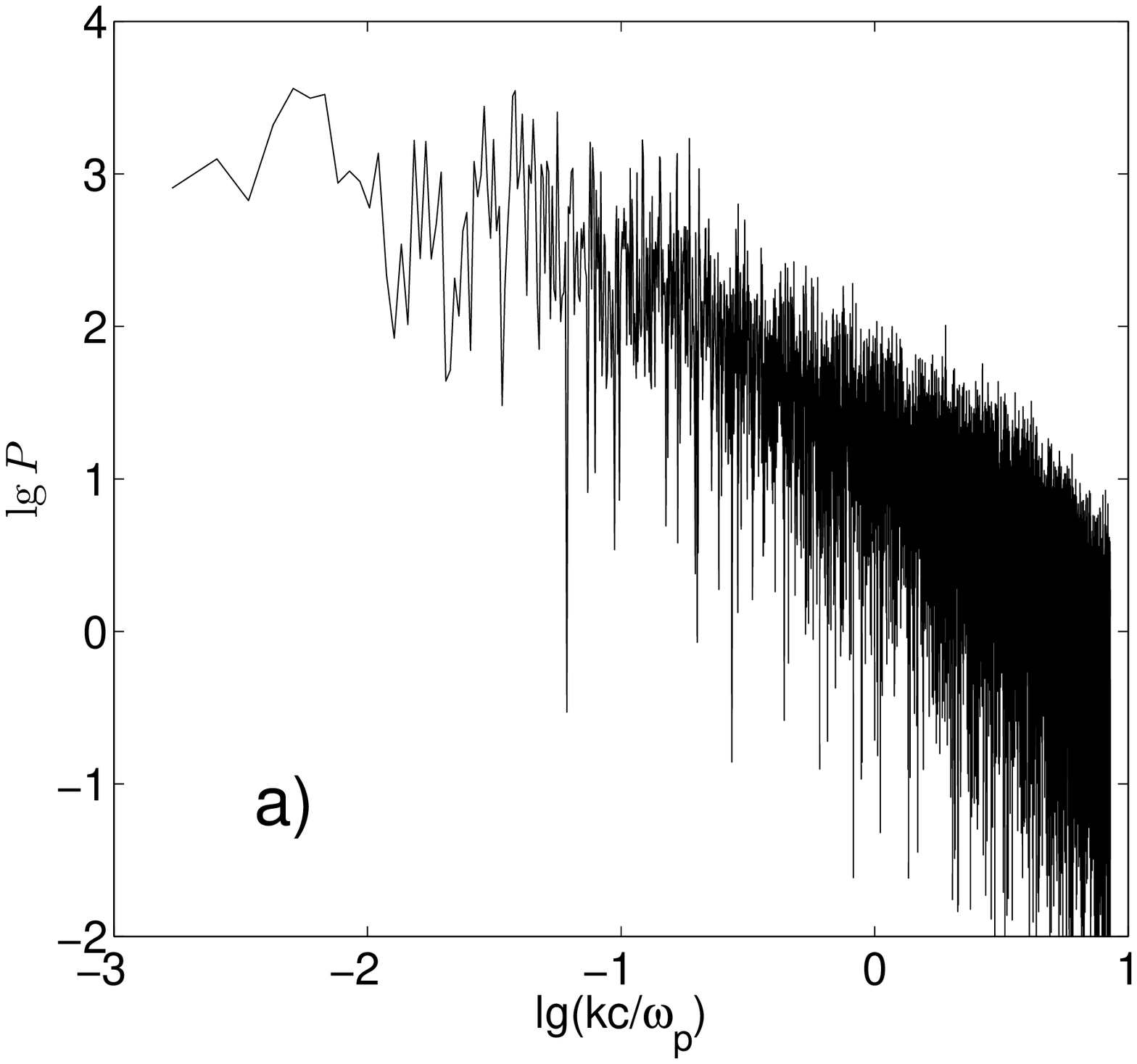}
\includegraphics[scale=0.4]{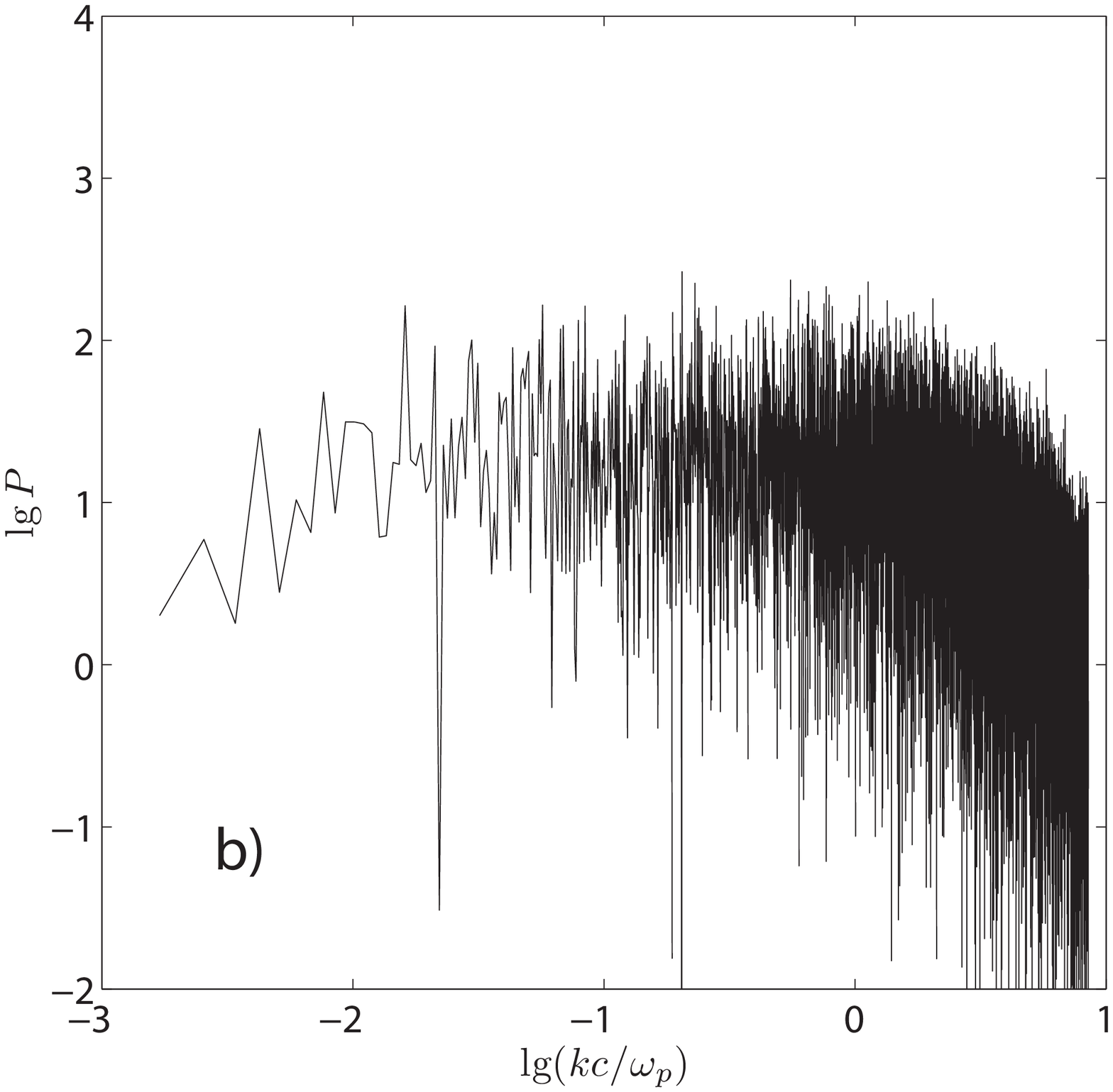}
\caption{The power spectrum of the fluctuations of the
longitudinal electric field in the region $0.25l_0<z<0.75l_0$
(a) and $1.5l_0<z<2l_0$ (b).}
\end{figure*}

\begin{figure*}
\includegraphics[width=14 cm,scale=0.5]{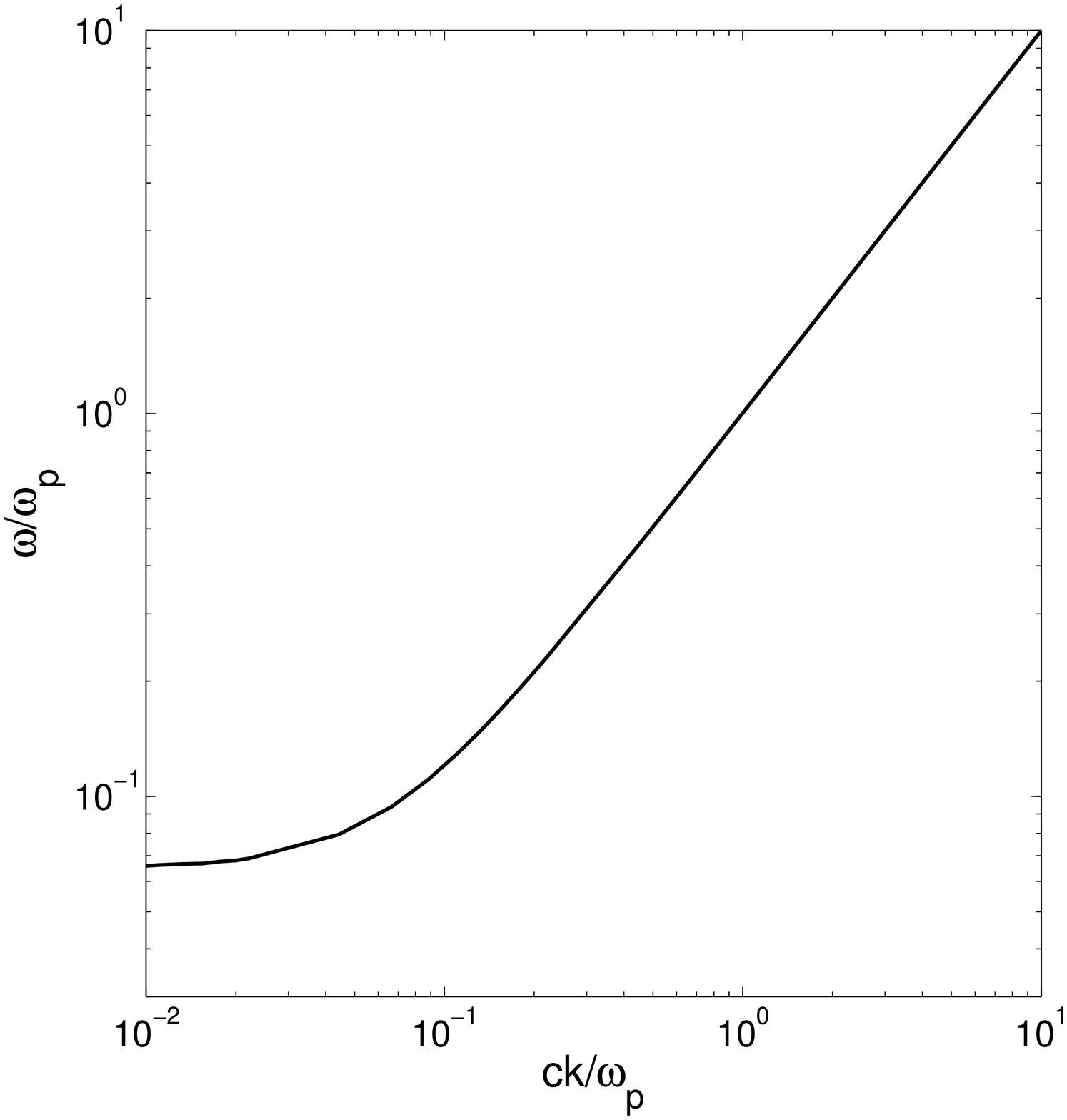}
\caption{The spectrum of the longitudinal oscillations in the plasma with the distribution functions of electrons and positrons shown in Fig. 1 by the thin solid and dotted lines, correspondingly.}
\end{figure*}

\section{Plasma oscillations excited in the course of the charge adjustment}

Inspection of Figs. 4 and 5 shows that strong fluctuations of
the longitudinal electric field are present in the region
$0<z<l_0$ where the background charge density varies. In Fig.
6, we show the power density spectrum of the fluctuations,
$P=(8\pi)^{-1}\vert E_{z,k}\vert^2$, in this region as well as
in the region $z>l_0$. One sees that in the last region, where
the background charge density is constant so that additional
redistribution of charges does not occur, only short-wavelength
fluctuations are present, which could be attributed to the
numerical noise. In the region $0<z<l_0$, the fluctuation
spectrum is extended to the long-wavelength domain, which means
that the charge redistribution is accompanied by generation of
long-wavelength plasma oscillations.

The longitudinal fluctuations in the
relativistic,one-dimensional plasma are described by the
dispersion relation (e.g. \citet{lyub96})
 \eqb
\omega_p^2\int\frac{f_+(p)+f_-(p)}{\gamma^3(\omega-kv)^2}dp=1;
 \label{disp}\eqe
where $f_{\pm}$ are the distribution functions of electrons and
positrons, correspondingly. The characteristic frequency of the
long wavelength oscillations,
 \eqb
\omega_0=\omega_p\left(\int\frac{(f_++f_-)dp}{\gamma^3}\right)^{1/2};
 \eqe
is determined by the low energy particles. In Fig. 7, we show
the numerical solution to the dispersion equation (\ref{disp})
for the distribution functions found in our simulations (and
shown in Fig. 1); one sees that $\omega_0\approx 0.06\omega_p$
in this case. According to Fig. 6a, the maximum of the
fluctuation energy, $kP$,  occurs at $k\sim 0.05\omega_p/c$,
which is close to $\omega_0/c$. In order to check this scaling,
we performed the same simulations with $\omega_p$ three times
larger than in the simulations shown in Figs. (2-5). In this
case, the maximum of the power spectrum shifted three times
towards a lower $k$, as one should expect.

Origin of these fluctuations is not very clear. For steady
state boundary conditions, one can easily construct a
completely time-independent state. The amount of particles
redirected at each point is determined by the spatial variation
of the Goldreich-Julian density. If the injection is
time-independent, the potential necessary to redirect the
required amount of particles is also time-independent. In
principle, the obtained state is unstable with respect to
two-stream instability because the momentum distribution of the
injected particles grows with $p$ at small $p>0$ whereas the
distribution of the redirected (and therefore having a negative
$p$) particles decreases with $p$ \citep{lyub_turb93}. However,
hardly ever this instability operates in the presented
simulations because numerical fluctuations of the electric
field are relatively strong so that the particles are
efficiently scattering smearing out any peculiarities on the
distribution function. By this reason, no signs of the
two-humped distribution function was seen in the simulations.

Fluctuations presumably arise because we inject equal amount of
electrons and positrons whereas in the true steady state, the
injected flux of electrons should be less by $(1/2)c\rho_0/e$
than the flux of positrons. This is because an electron
redirected at a point $z=z_1$ contributes twice into the charge
density at $z<z_1$ so that the total amount of injected
electrons should be less than that of positrons by amount of
the electrons redirected along the whole tube. If this
condition is not satisfied, extra charges (electrons in our
case) are expelled from the flow 
at the scale of the Debye length, which is determined by the
frequency $\omega_0$ for the injected plasma. Therefore a
"double layer" arises at the bottom of the tube; it is clearly
seen in Fig. 3 and less clearly (because of small scale) in
Figs. 4 and 5. Such a double layer is unsteady; it oscillates
at the time scale of $1/\omega_0$ thus slightly modulating the
injected plasma flow. This could give rise to the electric
field fluctuations in the tube because if the plasma flow is
somehow modulated, the decelerating electric field should
fluctuate in order to keep the number of redirected particles
fixed (the last is determined by the distribution of the
Goldreich-Julian charge density, which does not vary). On the
other hand, one has to stress that the modulation occurs at the
frequency $\omega_0$ for the injected plasma, which is less
than the frequency $\omega_0$ for the plasma in the tube
because the lowest energy particles in the injected plasma have
energy $\gamma\sim 10$ whereas in the tube, the particle
distribution is shifted to $\gamma\sim 1$. So the presented
interpretation is not straightforward.

In real pulsars, the characteristics of the injected plasma
depend on the conditions in the polar cap region therefore
there is no reason to expect that the polar cap accelerator
injects the particles in such a way that the difference in the
positron and electron fluxes is adjusted to the structure of
the magnetosphere in the light cylinder region  (recall that
$\rho_0$ is referred to the Goldreich-Julian density far from
the polar cap, somewhere at $\Omega r\sim 1$).  Moreover, the
plasma production process is by no means steady therefore one
can expect that the plasma flow in the open field line tube is
strongly modulated. Any modulation of the plasma flow would
result in fluctuations of the decelerating field along the tube
thus giving rise to long weavelength plasma oscillations in the
flow. Therefore we believe that such fluctuations are generic
in pulsar magnetospheres. Induced scattering of these
oscillations on the bulk plasma flow could give rise to high
brightness temperature pulsar radio emission
\citep{lyub_radio92,lyub_ind93,lyub96}.

\section{Adjustment of the electric current in the plasma flow}

One sees that when the system relaxes to the steady, on the
average, state, the current (and therefore the azimuthal
magnetic field) is established, which is determined by the
distribution of the Goldreich-Julian charge density. However,
there is no reason to expect that this current is consistent
with the current necessary to maintain the global structure of
the magnetosphere. As the condition $\mathbf{E\cdot B}=0$ is
fulfilled within the plasma flow, the structure of the  pulsar
magnetosphere satisfies the ideal MHD equations and as such do
not admit additional constraints on the current because the
current distribution in any MHD system is uniquely determined.
Specifically in the pulsar magnetospheres, the current
distribution is determined from the condition of the smooth
plasma transition through the light cylinder \citep{ingraham73,
contopoulos99,timokhin06}. Let us check what happens if the
current "demanded" by the global MHD solution differs from the
current established in the course of the charge density
adjustment, i.e., if the field (\ref{B0}) does not fit the MHD
structure of the magnetic field. For this purpose, we simulated
plasma flow within a tube of the length $l_0$ with a (generally
varying) magnetic field imposed at the upper edge of the tube.
As an initial condition, we used the distributions of the
fields and particles found at the end of the previous
simulations when the steady state was achieved in the region
$0<z<l_0$. We imposed the condition of free particle escape
from the upper edge of the tube and no particle inflow through
this edge. Conditions at the bottom edge remained the same as
in the previous simulations, i.e. the plasma is continuously
injected into the tube and $E_r(z=0)=0$. When we impose the
condition $B_{\varphi}(z=l_0)=B_{\varphi,0}$ at the upper edge
of the tube, nothing changes and the initial state persists
forever. However, when we slowly vary $B_{\varphi}(z=l_0)$ at
the upper boundary from $B_{\varphi,0}$ to another value,
$B_{\varphi,1}$, a discrepancy between the current in the
plasma and the imposed $B_{\varphi}$ results in the induction
electric field, which eventually establishes the necessary
current.

In the simulations, the magnetic field at the upper boundary
was changed according to the formula
 \eqb
B_{\varphi}(z=l_0)=B_{\varphi,1}+(B_{\varphi,0}-B_{\varphi,1})\exp\left(-t/\tau\right);
 \label{upperB}\eqe
In Figs. 8 and 9, we show how the system is relaxed to the
state with the azimuthal magnetic field
$B_{\varphi,1}=0.5B_{\varphi,0}$.  One sees that the electric
field arises near the upper edge of the tube so that positrons
are decelerated until some of them are redirected back thus
compensating the initial current. Eventually a steady state is
formed with the new magnetic field and with both electron and
positron backflows. The electrons are redirected along the
whole tube compensating variation of the background charge
density. The positron backflow is formed near the upper edge of
the tube.

Relaxation to the state with the current larger than the
initial one is achieved via formation of an enhanced electron
backflow, see Fig. 10.  In Fig. 11, we show the state with the
current having the sign opposite to that of the
Goldreich-Julian charge. Such a state is achieved by sending
more positrons back. So one concludes that the system could
easily maintain any current imposed at the outer edge, the
adjustment being achieved via formation of an appropriate
particle backflow.

\begin{figure*}
\includegraphics[width=14 cm,scale=1]{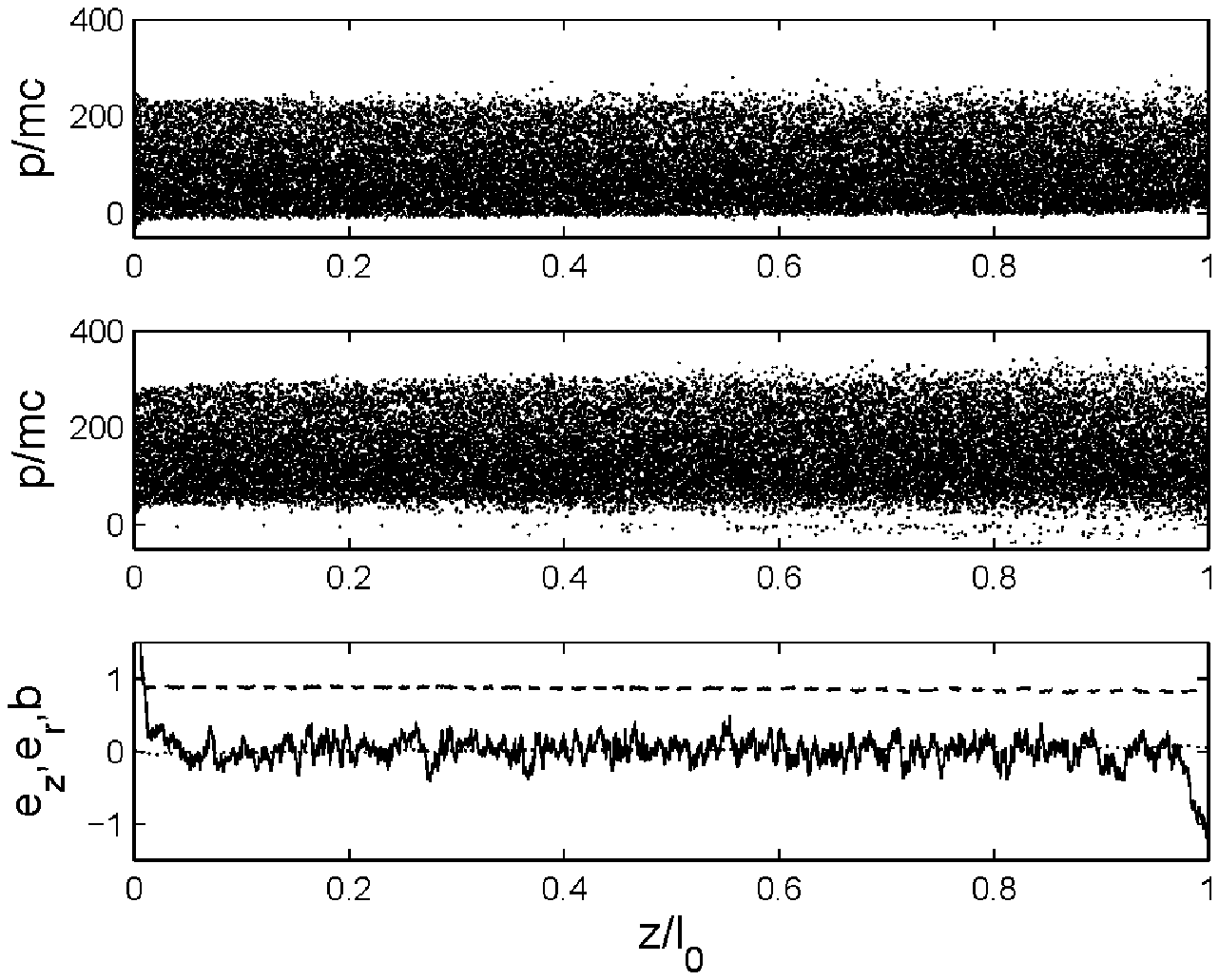}
\caption{Beginning of the current adjustment. Shown are the
phase spaces of the electrons (a) and positrons (b) and the
distribution of the normalized fields at $t=l_0/c$ after the
magnetic field at the upper (right) boundary of the tube begun
to vary according to Eq. (\ref{upperB}) with
$B_{\varphi,1}=0.5B_{\varphi,0}$ and $\tau=5l_0/c$. One sees
that the decelerating electric field is developed near the
upper edge and a weak positron backflow is formed; the magnetic
field is decreased a bit at this stage. }
\end{figure*}

\begin{figure*}
\includegraphics[width=14 cm,scale=1]{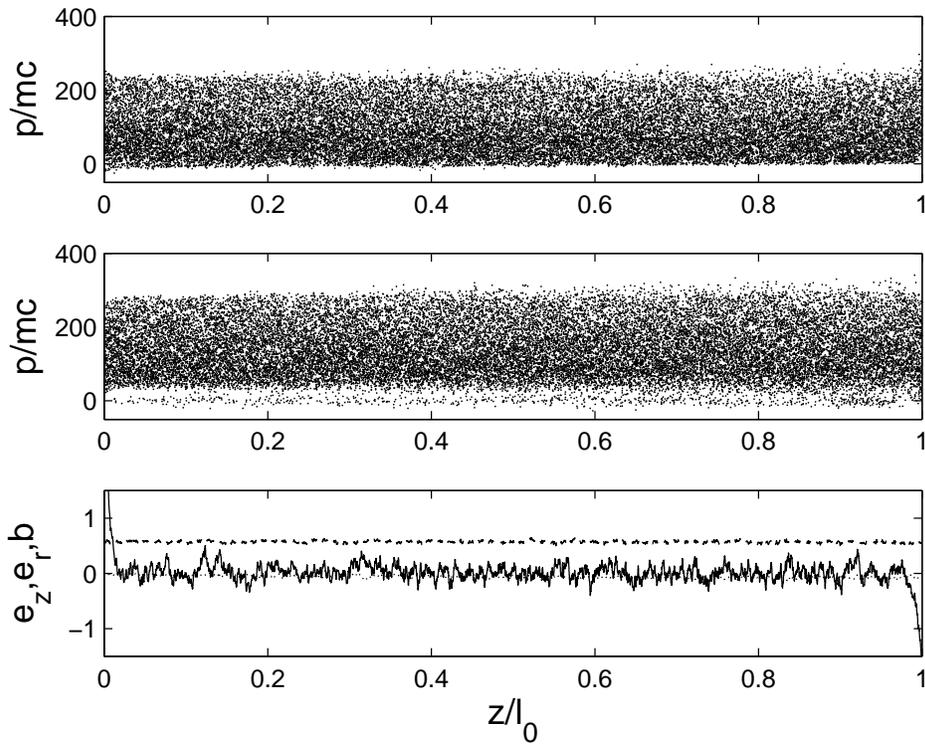}
\caption{The same as in Fig. 8 at $t=8l_0/c$. The necessary
positron backflow is already formed and the magnetic field is
nearly two times less than the initial one.}
\end{figure*}

\begin{figure*}
\includegraphics[width=14 cm,scale=1]{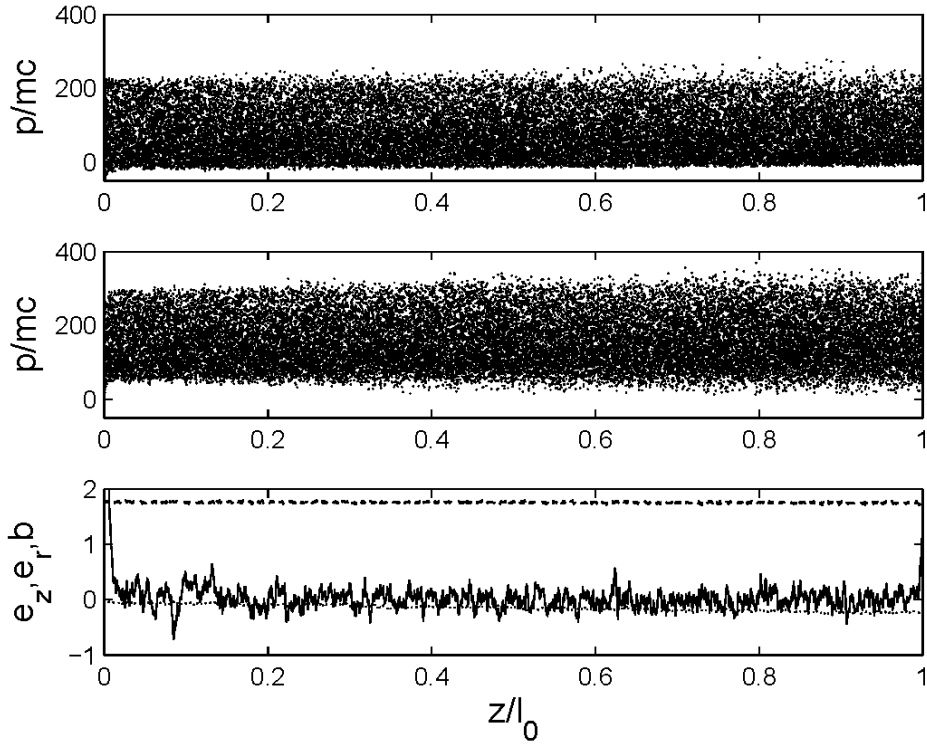}
\caption{The same as in Fig. 9 but for $B_{\varphi,1}=2B_{\varphi,0}$; $t=10l_0/c$; $\tau=5l_0/c$.}
\end{figure*}

\begin{figure*}
\includegraphics[width=14 cm,scale=1]{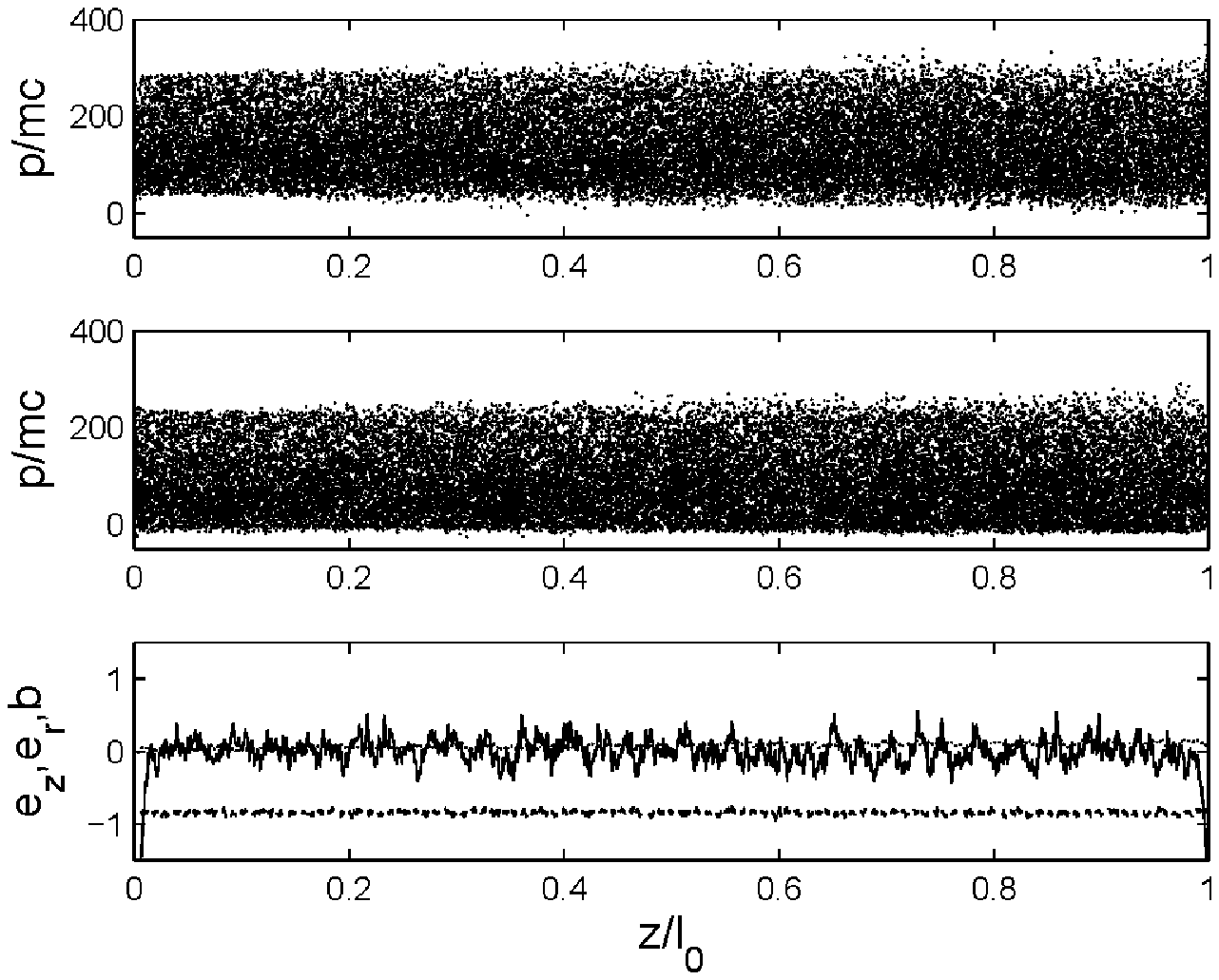}
\caption{The same as in Fig. 9 but for $B_{\varphi,1}=-B_{\varphi,0}$; $\tau=10l_0/c$; $t=30l_0/c$. }
\end{figure*}

\section{Conclusions}

We have demonstrated that the relativistic plasma flow in the
open field line tube of the pulsar easily screens the
rotationally induced longitudinal (along the magnetic field)
electric field and maintains any current demanded by the
boundary conditions.  This justifies the MHD approach to the
structure of the pulsar magnetosphere provided the plasma
density in the flow exceeds the Goldreich-Julian density. An
important point is that the adjustment of the electric charge
and the current is achieved via formation of a particle
backflow with the density roughly $\rho_{GJ}/e$. Depending on
the magnetospheric current and on the distribution of the
Goldreich-Julian charge density along the open field line tube,
the backflow could consist of electrons, positrons or both.

In these simulations, the pair plasma with the density
significantly larger than $\rho_{GJ}/e$ was continuously
injected from the bottom of the tube so that a steady state was
eventually achieved. In the pulsar magnetosphere, the plasma is
assumed to be produced at the bottom of the open field line
tube where primary particles are accelerated in the potential
gap (e.g., \citet{hibsch_arons01}). The present study shows
that one should expect a particle flux coming from above to the
gap. When the plasma fills the gap from above, the electric
field is shorted out, the particle acceleration is terminated
and the pair production ceases. Presumably this means that the
polar accelerator could not operate in the steady state regime.
One can expect some sort of spasmodic plasma production such
that the required current is maintained only on average. Future
models of the polar cap accelerator should take into account
the plasma flow onto the gap from the magnetosphere.

Another implication of this study is that the charge and
current adjustment is accompanied by continuous generation of
long-wavelength plasma oscillations. As a result of the induced
scattering of these oscillations in the bulk plasma flow, these
oscillations could be converted into the electromagnetic waves
with the frequency $\omega\sim\omega_p\sqrt{\gamma_{\rm max}}$,
where $\gamma_{\rm max}\sim 100$ is the Lorentz factor
corresponding to the maximum in the particle energy spectrum
\citep{lyub96}. The energy of these waves exceeds the energy of
the initial oscillations $\sim \omega_p\sqrt{\gamma_{\rm
max}}/\omega_0$ times. One can speculate that this process
gives rise to the observed pulsar radio emission (for more
details see \citet{lyub96}).

\acknowledgements I thank Saveli Rabinovich for help in
designing the code.  Useful discussions with Jon Arons and
Anatoly Spitkovsky are gratefully acknowledged. This work was
supported by the Israeli Science Foundation and by the
German-Israeli Foundation for Scientific Research and
Development.

\bibliographystyle{apj}

\bibliography{pulsar}

\end{document}